\documentclass[%
superscriptaddress,
 amsmath,amssymb,
 aps,
 prl,
 twocolumn,	 
floatfix,
citeautoscript
]{revtex4-2}
\usepackage[dvipdfmx]{graphicx}
\usepackage{amsthm}
\usepackage{amscd}
\usepackage{latexsym}
\usepackage{xcolor}
\usepackage{bm}
\usepackage{dcolumn}
\usepackage{hyperref}
\newlength{\figheight}
\def\la{\lambda}

\def\s{\sigma}
\def\g{\gamma}

\def\la{\lambda}
\def\La{\Lambda}
\def\Om{\Omega}

\makeatletter
\renewcommand{\appendix}{
\renewcommand{\thesection}{Appendix \Alph{section}\hspace{-.1em}}
\renewcommand{\thesubsection}
 {\Alph{section}\arabic{subsection}.\hspace{-.1em}}
\renewcommand{\thesubsubsection}
 {\Alph{section}\arabic{subsection}.\arabic{subsubsection}.\hspace
                                                            {-.1em}}
\@addtoreset{equation}{subsection}
\renewcommand{\theequation}{\Alph{section}.\arabic{equation}} 
\setcounter{section}{0}}
\makeatother

\begin{document}



\title{Exact real-time longitudinal correlation functions\\ of the massive XXZ chain}
\author{Constantin Babenko} 
\affiliation{Fakult\"at f\"ur Mathematik und Naturwissenschaften,
Bergische Universit\"at Wuppertal, 42097 Wuppertal, Germany}
\author{Frank G\"ohmann}
\affiliation{Fakult\"at f\"ur Mathematik und Naturwissenschaften,
Bergische Universit\"at Wuppertal, 42097 Wuppertal, Germany}
\author{Karol K. Kozlowski}
\affiliation{Univ Lyon, ENS de Lyon, Univ Claude Bernard, CNRS,
Laboratoire de Physique, F-69342 Lyon, France}
\author{Jesko Sirker}
\affiliation{Department of Physics and Astronomy, and Manitoba Quantum Institute,
University of Manitoba,
Winnipeg R3T 2N2,
Canada}
\author{Junji Suzuki}
\affiliation{Department of Physics, Faculty of Science, Shizuoka University, Ohya 836, Suruga, Shizuoka, Japan}

\begin{abstract}
We apply a recently developed thermal form factor expansion method to
evaluate the real-time longitudinal spin-spin correlation functions of
the spin-$\frac{1}{2}$ XXZ chain in the antiferromagnetically ordered
regime at zero temperature. An analytical result incorporating all types
of excitations in the model is obtained, without any approximations.
This allows for the accurate calculation of the real-time correlations
in this strongly interacting quantum system for arbitrary distances
and times.
\end{abstract}
\maketitle


Dynamical correlation functions relate experimental observables like
structure factors and transport coefficients to the microscopic
details of quantum many-body systems. They are notoriously hard
to calculate as they simultaneously probe all time and length
scales. A particular class of many-body systems are integrable
one-dimensional (1d) systems with short-range interactions. Due
to the existence of a large number of local conserved quantities,
they exhibit a peculiar phenomenology: they do not relax to a
thermal equilibrium and can possess spin and charge currents
which do not fully decay in time. These unusual properties
are not only of fundamental interest but can be observed in
experiments on systems which are almost integrable. This includes
realizations of the 1d Bose gas in which quench dynamics as well
as dynamical correlations in equilibrium have been studied
\cite{par04,HofferberthLesanovsky,KinoshitaWenger,GringKuhnert,GuarreraMuth}. 
The Heisenberg chain can be realized using cold atomic gases in
optical lattices giving direct access to the spin dynamics in the
spatio-temporal domain and to spin transport phenomena
\cite{FukuharaKantian,FukuharaSchauss,HildFukuhara, JepsenGrillDimitrovaHoDemlerKetterle}. 
Good realizations of the Heisenberg chain also occur as sub-structures
in solid state systems
\cite{MotoyamaEisaki,TakigawaMotoyama,ThurberHunt}. Here they provide access 
to response functions in the momentum-frequency domain as, for example, the
dynamical spin-structure factor (DSF) \cite{MourigalEnderle}. They also allow
for the direct measurement of transport coefficients
\cite{SologubenkoGianno,SologubenkoBerggold,HlubekRibeiro,KohamaSologubenko}. 
Recent attempts to devise a general description of 1d
close-to-integrable systems resulted in interesting phenomenological
theories like the non-linear Luttinger liquid 
\cite{ImambekovGlazman,GlazmanReview,PereiraWhite,PereiraIJMP,SirkerLL} 
or generalized hydrodynamics
\cite{AlvaredoDoyon,UrichukOez,BulchandaniVasseur,BertiniCollura,deNardisBernardDoyon}. 
It is highly desirable to underpin such new phenomenologies with
microscopic calculations. We note, moreover, that dynamical correlation
functions in cold atomic gases can now be tested at time scales which
are beyond the reach of modern numerical techniques. An exact
calculation of dynamical correlation functions of 1d integrable models
is therefore important for our understanding of state-of-the-art
experiments. At the same time, it provides benchmarks for phenomenological
theories and numerical methods.

Exact results on correlation functions are rather rare, even in low
dimensions, and are mostly related to models belonging to the Free
Fermion (FF) category \cite{LSM61}. For the Ising model in the scaling
limit, a remarkable link to the Painlev{\'e} equations was established
\cite{MTW76}. Dynamical correlations were studied for the XY model
\cite{Niemeijer67} and interesting phenomena of thermalization were
addressed.
An important next step was to go beyond FF and deal exactly
with interacting systems.
The vertex operator approach (VOA) opened up a new avenue to do so
based on form factor expansions
\cite{JMMN92}. The evaluation of the two- and four-spinon
contributions to the transverse DSF of the massive XXZ chain
\cite{BKM98,AbadaBougourzi, CauxMosselCastillo} and of the four-spinon contribution
in the XXX limit \cite{CaHa06} were important outcomes of this
method. The complexity of the resultant multiple integrals has,
however, hindered any further analysis.
A hidden FF structure in the XXZ model was unveiled in
Refs.~\cite{BJMST08a,JMS08}, and its remarkable outcome, the
fermionic basis, yields exact correlation functions up to
considerably large distances \cite{SABGKTT11,MiSm19}. The
application of this method, however,  has been limited to the
static case so far.
The quantum inverse scattering method (QISM) provides a complementary
approach \cite{KBIBo,KMT99a}, based on a determinant
formula for the scalar product of on- and off-shell Bethe
vectors~\cite{Slavnov89}. Under restrictions on the possible excitations,
it successfully reproduces the asymptotic correlation functions
\cite{KKMST11b, KKMST12} predicted by conformal field theory (CFT) 
\cite{Cardy86}. Yet, in order to recover the DSF for the full range of
frequencies and momenta, bound states must be taken into account,
which are neglected in the CFT limit \cite{CaMa05}.

In this Letter we employ a recently developed \cite{GKKKS17,BGKS20app}
form factor expansion for real-time correlation functions
in equilibrium and obtain a simple, explicit, closed-form
expression including all orders of excitations. We consider the
XXZ chain with Hamiltonian
\begin{equation}\label{hamiltonian}
    H \!= \!J \sum_{j = 1}^L \Bigl\{ \sigma_{j-1}^x  \sigma_j^x +  \sigma_{j-1}^y  \sigma_j^y
                 + \Delta  \sigma_{j-1}^z  \sigma_j^z \Bigr\}
		 - \frac{h}{2} \sum_{j=1}^L  \sigma_j^z
\end{equation}
for length $L\to\infty$. Here $\sigma^\alpha_j$ are Pauli matrices, and $J>0$
is the exchange constant. We restrict ourselves to the
antiferromagnetically ordered regime, characterized by values
$\Delta=\cosh(\gamma )>1$ of the anisotropy and by magnetic fields
in the range $0<h< 4J \sinh (\gamma )\vartheta^2_4(0|q)$. Here we
have set $q={\rm e}^{-\gamma}$, and the $\vartheta_a$ denote elliptic
theta functions \cite{dlmf}. The special functions appearing here
and below are summarized in the Suppl.~Mat.~\cite{SupplMaterial}.

In the antiferromagnetically ordered regime, the one-body properties
of the elementary excitations are characterized by
\begin{align} p(\theta) & =
\frac{\pi}{2} + \theta -i \ln \biggl( \frac{\vartheta_4 (\theta + i
\gamma/2| q^2)}{\vartheta_4 (\theta - i \gamma/2| q^2)} \biggr),  \label{dressede} \\
 \varepsilon(\theta) & = \frac{h}{2} - 2 J \sinh(\gamma) \vartheta_3(0|q) \vartheta_4(0|q) \frac{\vartheta_3
(\theta|q)}{\vartheta_4 (\theta|q)} ,
\end{align}
where $p(\theta)$ is the dressed momentum, $\varepsilon(\theta)$
the dressed energy and $\theta$ the rapidity of the quasiparticle.
The interaction between excitations is described by
the soliton scattering matrix,
 \begin{equation}\label{scattering_phase}
 S(\theta)= {\rm e}^{i ( \frac {\pi}{2}+ \theta)}
\frac{\Gamma_{q^4} \bigl(1 + \frac{i \theta}{2\gamma}\bigr)
	                     \Gamma_{q^4} \bigl(\frac{1}{2} - \frac{i \theta}{2\gamma}\bigr)}
		            {\Gamma_{q^4} \bigl(1 - \frac{i \theta}{2\gamma}\bigr)
			     \Gamma_{q^4} \bigl(\frac{1}{2} + \frac{i \theta}{2\gamma}\bigr)} ,
 \end{equation}
where $\Gamma_q$ denotes the $q$-gamma function. Due to the
Yang-Baxter integrability of the model, any multi-particle scattering
can be reduced to multiple two-body scattering events. One then
naturally expects that correlation functions can be described solely
by (\ref{dressede})-(\ref{scattering_phase}). We will show that an
all-order expansion of the longitudinal dynamical correlation
functions can indeed be described by these physical quantities with
supplemental special functions of the $q$-gamma function family.

Our framework combines the QISM and the quantum transfer matrix (QTM) method
\cite{Suzuki85,SAW90,Kluemper92}. The latter has been devised for the
investigation of finite temperature bulk quantities and static
correlation functions
\cite{GKS04a,BGKS07}. 
 It was generalized in \cite{GKKKS17}, inspired by
\cite{Sakai07}, to obtain form factor expansions of dynamical correlations
at  finite temperatures. We thus call it the thermal form factor
expansion method.  It has 
been successfully applied to the
analysis of a FF model, the XX model \cite{GKKKS17, GKSS19}. 

Although we are interested in the dynamical correlations in the ground
state, we start from finite temperatures and consider the limits $h, T
\rightarrow 0$. This may look redundant at first, but there are
advantages of this approach.
It is well known that the diagonalization of the Hamiltonian leads to
 string excitations of various lengths \cite{ BVV83, ViWo84, Takahashi99}.
  These are solutions of the Bethe ansatz
equations which form regular patterns (`strings') in the complex plane
for $L \rightarrow \infty$. Some quantities which characterize the
correlations become singular if they are evaluated at the ideal string
positions, e.g., $S(\pm i\gamma)$.  The variety of string excitations
and singularities leads to serious technical difficulties. The VOA
provides an alternative description, free from string excitations, but its answer suffers from the high intricacy of multiple integrals.
On the other hand, the QTM method, based on a mapping of the 1d
quantum system to a two-dimensional classical system, does not
directly deal with the Hamiltonian but rather with a transfer matrix
acting in an auxiliary space. The possible excitations are thus
different from those in the Hamiltonian basis.  A previous study, using the
higher level Bethe ansatz equations, concludes that only simple
excitations are possible for $L \rightarrow \infty$ and $h, T
\rightarrow 0$ with the limit $T\to 0$ taken first \cite{DGKS15b}.
Their distribution in the complex rapidity plane can be interpreted as
particle-hole excitations. A rapidity $y_j$ of a particle excitation
is situated on a curve located above $[-\pi/2; \pi/2]$ such that ${\rm
Im} \,y_j \sim \frac{\gamma}{2}$, while a hole rapidity $x_j$ is
located below and ${\rm Im}\, x_j \sim -\frac{\gamma}{2}$.  These
excitations are \textit{not} 2 strings since $\Re(y_j - x_{l})$ is
generically non-zero and $\Im( y_j - x_{l})$ does \textit{not}
approach $\gamma$, even for $L\rightarrow \infty$.  Thus, the QTM
excitations do not produce the aforementioned singularities.

For the longitudinal correlation functions 
\begin{equation}
\langle
\sigma^z_1(t)\sigma^z_{m+1}(0)\rangle =\lim_{T\to 0}
\mbox{tr}\,\{\sigma^z_1(t)\sigma^z_{m+1}(0)\exp(-H/T)\}/Z \nonumber
\end{equation} 
where $Z$ is the partition function, the relevant excited states
consist of an equal number of particles and holes. Thus, the resultant
form factor expansion involves a sum over $\ell$, the number of
particles and holes, and a sum over their possible locations. The
higher level Bethe ansatz analysis shows that they  obey a one-body equation
  for $T \rightarrow 0$.  
 The sum over the possible locations  is then replaced by simple integrations \cite{SupplMaterial},
\begin{align} 
& G(m,t):=\langle \sigma_1^z (t) \sigma_{m+1}^z(0) \rangle  -
 (-1)^m \Bigl(\frac{\vartheta_1'(0|q)}{\vartheta_2(0|q) } \Bigr)^2  \nonumber \\
   \! &=\!
\sum_{\substack{\ell \ge 1\\k = 0, 1}} \!\frac{(-1)^{km}}{(\ell!)^2}
\!\int_{C_-}
\!\frac{d^{\ell } x}{(2\pi )^{\ell }}  
 \! \int_{C_+}
\!\frac{d^{\ell } y}{(2\pi )^{\ell }}  
      {\rm e}^{ - i \sum_{j=1}^{\ell } (  m p(x_j)-\varepsilon(x_j) t)} \nonumber  \\
     &\phantom{ccc}\times  {\rm e}^{  i \sum_{j=1}^{\ell }  (  m p(y_j)- \varepsilon(y_j) t )}
     {\cal A}^{zz} (\{x_i\}_{i=1}^{\ell },  \{y_j\}_{j=1}^{\ell }|k)  \label{ffexpansion} \\
    & =\sum_{\ell \ge 1} I_{\ell }(m,t).  \nonumber
\end{align}
The integer $k \in \{0, 1\}$ labels the degenerate ground states and
we have subtracted the contribution of the staggered magnetization.
There is some freedom to choose the contours $C_{\pm}$: the
simplest choice is to take straight segments of length $\pi$ whose
imaginary parts are $\pm \gamma/2 +\delta$ where $\delta$ is
positive. We will discuss the optimal choice of the contours for a
numerical evaluation later.  


The main purpose of the present report is to present the explicit form
of ${\cal A}^{zz}$. 
It consists of determinants of two ${\ell }\times{\ell }$ matrices $\mathcal{M}$ and $\hat{\mathcal{M}}$ and a scalar part.
 For a compact presentation, we will use the shorthand notations
\[
P_j={\rm e}^{2i y_j}, \quad H_j= {\rm e}^{2 i x_j}, \qquad 1 \le j \le \ell, 
\]
and introduce the basic hypergeometric series \cite{dlmf,  GaRa04},
\begin{align}
&\Phi_1(P_j) = 
\phantom{}_{2\ell} \Phi_{2\ell-1}
\left (
{q^{-2},      \{ q^2  \frac{P_j} {P_{i}} \}_{i \ne j},          \{  \frac{P_j} {H_{i}} \}_{i=1}^{\ell } \atop
  \phantom{--}   \{   \frac{P_j} {P_{i}} \}_{i \ne j},          \{  q^2\frac{P_j} {H_{i}} \}_{i=1}^{\ell }       }
;q^4, q^4 
\right),     \nonumber \\ 
&\Phi_2(P_j, P_i ) =\nonumber \\
&\phantom{}_{2\ell } \Phi_{2\ell -1}
\left (
{\! q^{6},         \{ q^6  \frac{P_i} {P_{r}} \}_{r \ne i,j},    q^2  \frac{P_i} {P_j} ,       \{  q^4\frac{P_i} {H_{r}} \}_{r=1}^{\ell}  \atop
   \{ q^4  \frac{P_i} {P_{r}} \}_{r \ne i,j},   q^8   \frac{P_i} {P_j},       \{  q^6\frac{P_i} {H_{r}} \}_{r=1}^{\ell}   }
;q^4, q^4 \!
\right).   \label{defPhi2}  
\end{align}
They originate from sums of residues of the soliton $S$ matrix at a particular
series of poles. 
We further introduce
\begin{multline}\label{defrn}
r_{\ell}(P_j,P_i) \\
= \frac{ q^2(1-q^2)^2\frac{P_i}{P_j}}{ (1-\frac{P_i}{P_j})(1-q^4 \frac{P_i}{P_j} )      }
\prod_{r=1}^{\ell} \frac{ 1- \frac{P_i}{H_{r}} }{ 1- q^2\frac{P_i}{H_{r}} }  
\prod_{r \ne i,j}^{\ell} \frac{ 1- q^2\frac{P_i}{P_{r}} }{ 1- \frac{P_i}{P_{r}} } 
\end{multline}
and conveniently write $\Psi_2(P_j, P_i)= r_{\ell}(P_j,P_i) \Phi_2(P_j, P_i)$.
The matrix element  $\mathcal{M}_{ij}$ is then given by
\begin{equation}\label{defPhi}
\mathcal{M}_{ij}=\delta_{ij}D_{ij}+ (1-\delta_{ij})E_{ij}
\end{equation}
with
\begin{align}
D_{ij}&=  \overline{\Phi}_1(P_j) -{\Phi}_1(P_j)  (-1)^k \prod_{r=1}^{\ell} \frac{S (y_j- y_{r})}{S (y_j- x_{r})} ,  \label{defD}\\
E_{ij}&= - \overline{\Psi}_2(P_j,P_i) + {\Psi}_2(P_j,P_i)   (-1)^k  \prod_{r=1}^{\ell} \frac {S (y_i- y_{r})}{S (y_i- x_{r})}.  \nonumber 
\end{align}
Here we define for any function $g(P_1, \cdots, H_1, \cdots )$,
\[
{\bar g}(P_1, \cdots, H_1, \cdots) :=g(P_1^{-1},\cdots, H_1^{-1}, \cdots).
\]
The matrix element $\hat{\mathcal{M}}_{ij}$ is obtained from $\mathcal{M}_{ij}$ by
replacing all $y_r \leftrightarrow -x_r$. 
Then ${\cal A}^{zz}$ is  explicitly represented as
\begin{multline}\label{Azz_formula}
 {\cal A}^{zz}  =
   {\rm det}(\mathcal{M})  \,{\rm det}(\hat{\mathcal{M}}) \Bigl (  \frac{ \mu^{\ell}  \vartheta'_1(0|q)   \sin {\cal P} }{ \vartheta_1(\Sigma|q)    } \Bigr)^2  \\
\times  \frac{   \prod_{1\le i< j \le {\ell}}  \psi_D(x_i-x_j)  \psi_D(y_i-y_j)}{\prod_{i,j} \psi_D(x_i-y_j)}.
 \end{multline}
We set  ${\cal P}= \frac{\pi k}{2}+ \sum_{l} \frac{p(y_l)-p(x_l)}{2}$, 
$\Sigma=-\frac{\pi k}{2}+ \sum_{l} \frac{y_l-x_l}{2}$,
\begin{equation} \label{defpsi}
   \psi(\theta) = \Gamma_{q^4} \bigl(\frac{1}{2} - \frac{i \theta }{2\gamma}\bigr)
               \Gamma_{q^4} \bigl(1 - \frac{i \theta}{2\gamma}\bigr)
               \frac{G^2_{q^4} \bigl(1 - \frac{i\theta}{2\gamma}\bigr)}
                    {G^2_{q^4} \bigl(\frac{1}{2} - \frac{i \theta}{2\gamma}\bigr)},
\end{equation}
and $\psi_D(\theta)= \vartheta^2_1(\theta|q^2)\psi(\theta)\psi(-\theta)$.
The symbol   $G_q$ stands for  the $q$-Barnes' $G$ function.
The overall constant $\mu$ is given by $ \vartheta'_1(0|q^2) \psi(0)$.

We stress again that the compact formula (\ref{Azz_formula}) is valid
for arbitrary ${\ell}$ and is free from any approximations.
A similar all order formula has been derived for a quantum field
theoretical model in Ref.~\cite{FSmirnov86} but so far not for lattice
models.
As an analytic benchmark, we can show that Eq.~(\ref{Azz_formula}) for
$\ell = 1$ successfully reproduces \cite{BGKS20app} the result of the
VOA for two spinons
\cite{Lashkevich02}. Generally, there is a conjecture
\cite{DGKS16b} about the equivalence of the
contributions from $2{\ell}$-spinons and from ${\ell}$-particle/hole
excitations, which will be analyzed in detail in a separate
publication.

\begin{figure}[!h]
 \centering
 \includegraphics[width=0.75\columnwidth]{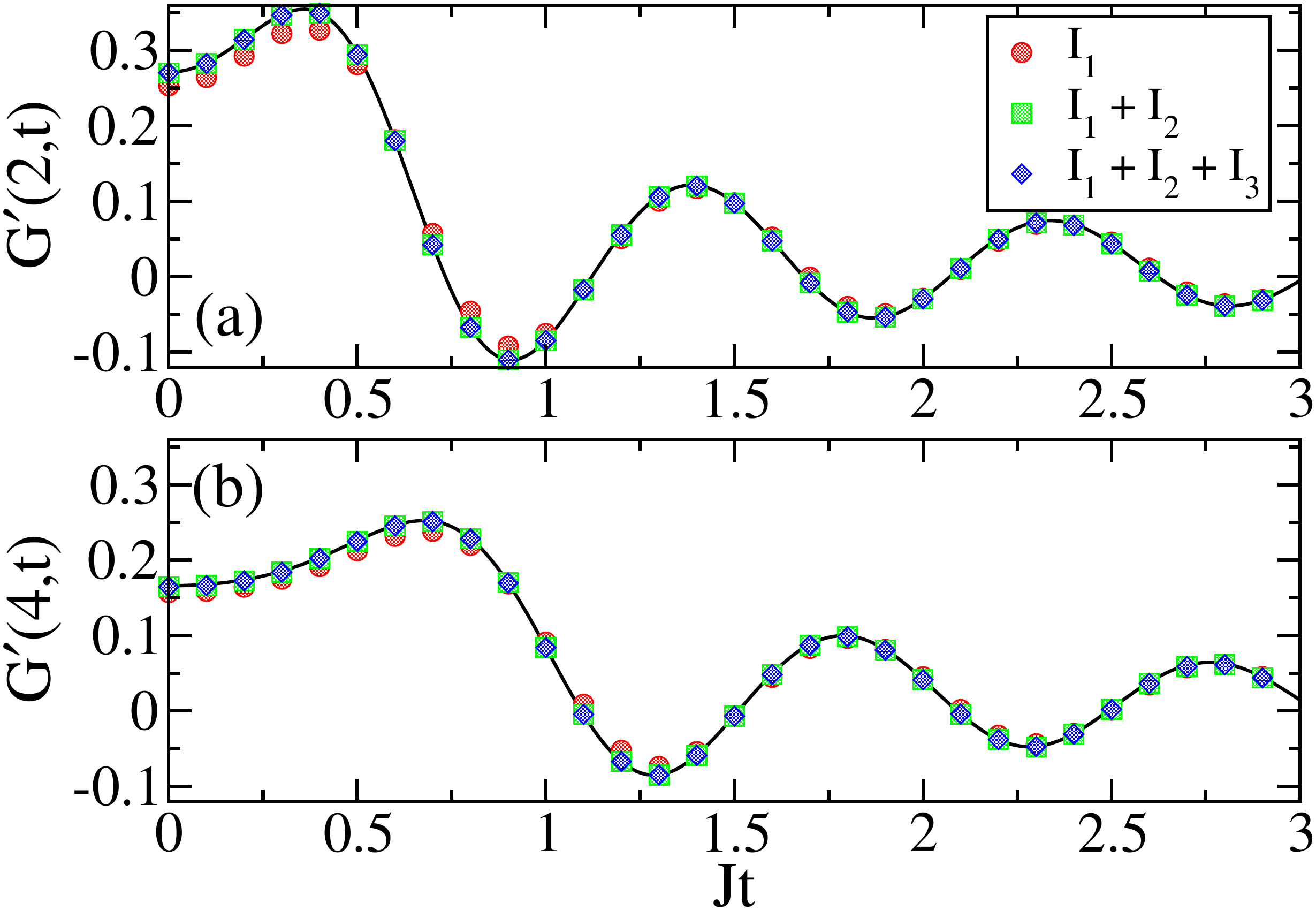}
 \caption{The real part $G'(m,t)\equiv \Re G(m,t)$  for $\Delta=1.2$ with (a) $m=2$, and (b) $m=4$:
 Contributions of a thermal form factor expansion
 (symbols) are compared to tDMRG data (lines).}
 \label{fig:m12D12}
 \end{figure}

Our main result (\ref{Azz_formula}) is very efficient for the numerical
evaluation of the real-time dynamics. It allows us to obtain $G(m,t)$
for {\it arbitrary} distances and times, thus going far beyond of what
can be achieved by purely numerical algorithms.
In Ref.~\cite{DGKS16b} the static correlation functions were 
investigated within the same framework, but using a Fredholm determinant
representation for ${\cal A}^{zz}$.
This inevitably included a numerical discretization approximation
\cite{Bornemann10}. Although the results seem highly precise, it is
important to check them independently as the actual evaluation
involves numerical integrations. In the static case, we can establish
results with high precision by a relatively small number of sampling
points $n_p$ for $2{\ell} $ multiple integrations, see
Table~\ref{tab:static123phm2}. 
\begin{table}[!h]
\begin{center}
\begin{tabular}{|l|c|c|c|c|c|}
\hline
$\Delta$ & $1.1$&   $1.3$ \\
\hline
$I_1 (2,0)$ & $0.2297348$&   $0.2357141$  \\
$I_2 (2,0)$ & $3.913377\times 10^{-2}$ &    $6.978269\times 10^{-3}$   \\
$I_3 (2,0)$ &  $1.614912\times 10^{-3}   $&   $2.120959\times 10^{-5}$   \\
\hline
$(I_1+I_2+I_3)$/exact& 0.99951&    0.999998  \\
\hline
\end{tabular}
\caption{Contributions to the static correlation $G(2,0)$ from small ${\ell}$ excitations
for $\Delta=1.1,\, 1.3$. The last row shows the ratios of the sums of
the first three terms in Eq.~\eqref{ffexpansion} and the exact values
\cite{TKS04}, demonstrating that keeping only excitations with $\ell\leq 3$ already leads to highly accurate results.}
\label{tab:static123phm2}
\end{center}
\end{table}
Once the system starts to evolve in time, however, the numerical
difficulty rapidly increases and it becomes necessary to replace the
numerical estimation of the Fredholm determinants by an analytic one.
To assess the accuracy of a truncated thermal form factor expansion in
the dynamic case, we compare results based on the $\ell\leq 3$
contributions with data obtained by a time-dependent density matrix
renormalization group algorithm (tDMRG) \cite{EnssSirker} in
Fig.~\ref{fig:m12D12}.
The plots clearly indicate the importance of including not just the
$\ell=1$ but also at least the $\ell=2$ contribution to obtain results
which agree with the tDMRG data on this scale. 
%
Based on this comparison, we restrict the following numerics to
excitations up to $\ell=2$ and leave a more detailed discussion of the
contributions of higher order excitations for a future study.

\begin{figure}[hbtp] \centering 
  \includegraphics[angle=90,width=0.75\columnwidth]{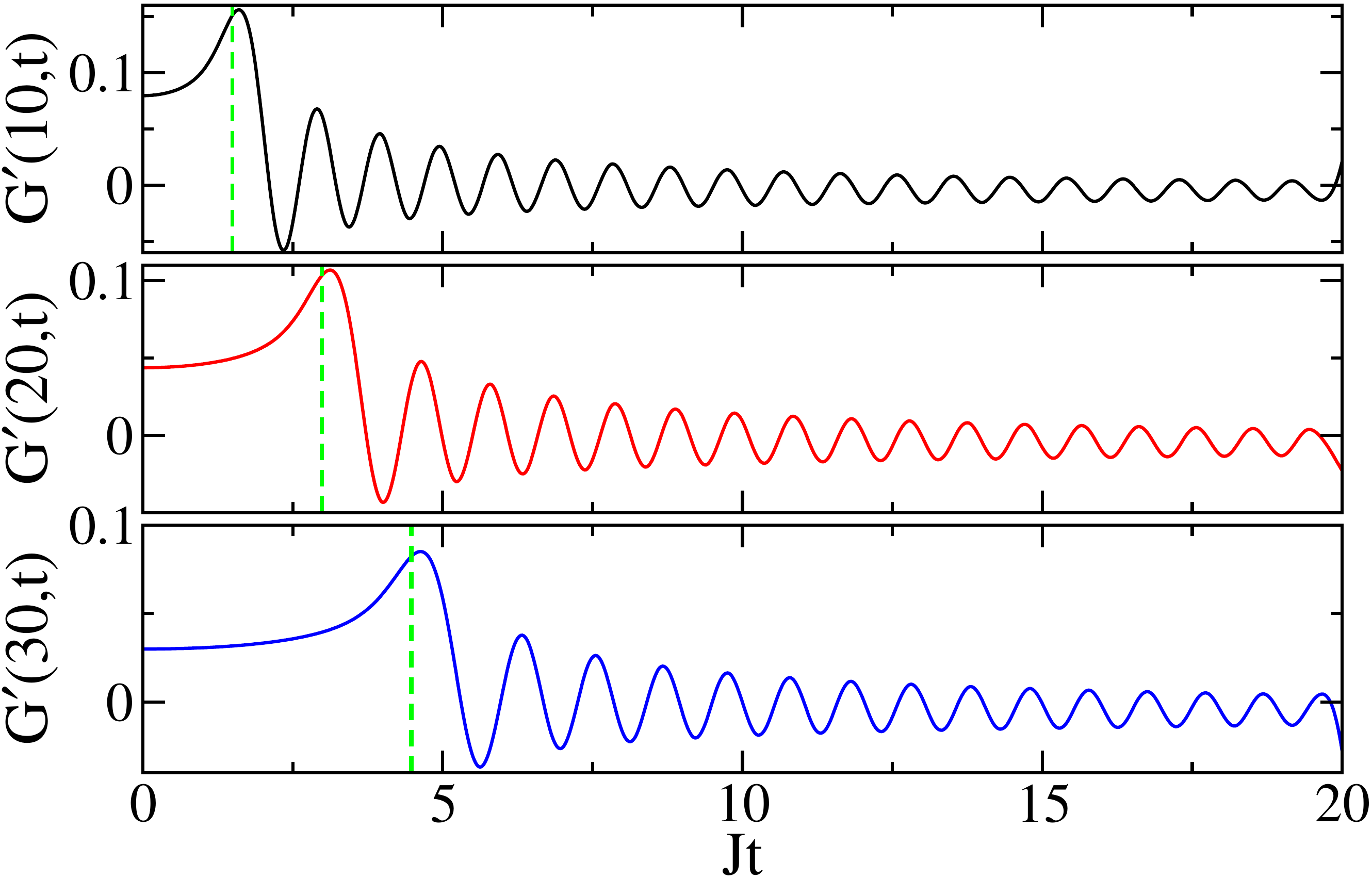}
\caption{The real part $G'(m,t)$ for different $m$ and $\Delta=1.2$. 
The dashed lines indicate the critical times $Jt_{c_1}$ and
$Jt_{c_2}$. For $\Delta \rightarrow 1$, $t_{c_1}\rightarrow t_{c_2}$
so there exists only an extremely narrow precursor regime (not visible
on this scale) for $\Delta=1.2$.}
\label{fig:m102030} 
\end{figure}

There are three asymptotic regimes, characterized by two critical
velocities $v_{c_1}< v_{c_2}$
\cite{DGKS16a}, or critical times $t_{c_a}=
m/v_{c_a}\,(a=1,2)$. We talk of the space regime if $t< t_{c_2}$, the
precursor regime if $t_{c_2}<t<t_{c_1}$, and the time regime if
$t>t_{c_1}$. The qualitative differences between the regimes can be
better seen for larger $m$. This is immediately handled by the form
factor expansion approach since $m$ enters as a mere
parameter. 
The correlation functions stay largely flat in the space regime, see
Fig.~\ref{fig:m102030}. Towards the edge of the regime, there occurs
an enhancement. After a transient behavior in the precursor regime,
the correlation exhibits an oscillatory behavior.
%

%

Let us now briefly discuss some technical issues in evaluating
Eq.~(\ref{ffexpansion}). In the time regime, the phase factors can
lead to serious numerical instabilities. The ordinary strategy to
overcome this problem is to deform the integration contours, making
them locally identical to the steepest descendent paths (SDP). This,
however, does not work naively in the present case, since the SDPs for
particles and holes intersect which leads to kinematic poles due to 
$\psi_D(y_j-x_i)$ in the denominator of Eq.~(\ref{Azz_formula}).
To solve this problem, we take advantage of the QTM formulation: we
return to finite temperatures and rewrite the formula in such a way
that the contribution from the intersecting part is multiplied by the
exponentially small factor $ {\rm e}^{-1/T}$. Thus, in the zero
temperature limit, one can neglect contributions from the kinematic
poles. As a result, the two contours become disentangled and can be
treated separately.
%

%
\begin{figure}[!ht]  \centering 
 \includegraphics[width=0.75\columnwidth]{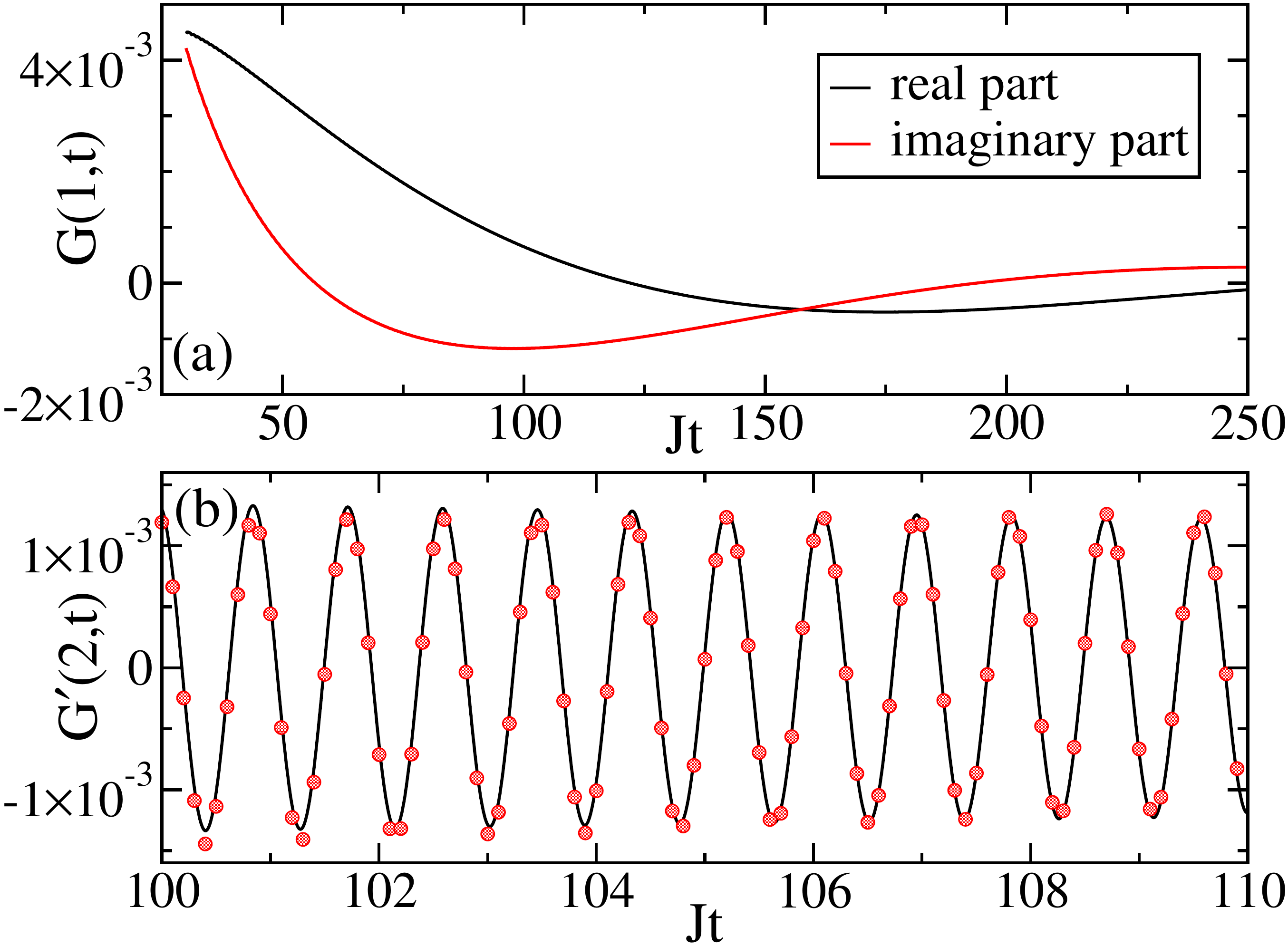} 
 \caption{(a) $G(1,t)$ at long times for $\Delta=1.2$. (b) Comparison
 of $G'(2,t)$ obtained by using the form factor expansion (symbols)
 with the two-spinon asymptotics (line) for $\Delta=1.4$.}
 \label{fig:Larger_time} \end{figure}
%
Thanks to this trick, stable calculations at long times become
possible, see Fig.~\ref{fig:Larger_time}(a). As a further test in the
time regime, we compare in Fig.~\ref{fig:Larger_time}(b) our results
to those obtained from the two-spinon term in a saddle point
approximation
\cite{DGKS16a}, which is expected to be
valid asymptotically in time. The predicted asymptotic behavior is
given by $G(m,t) \sim {\rm e}^{i \omega t}/t $ for $m$ even with
$\omega\sim J$.

 
 \begin{figure}[hbtp] 
\centering 
 \includegraphics[width=0.75\columnwidth]{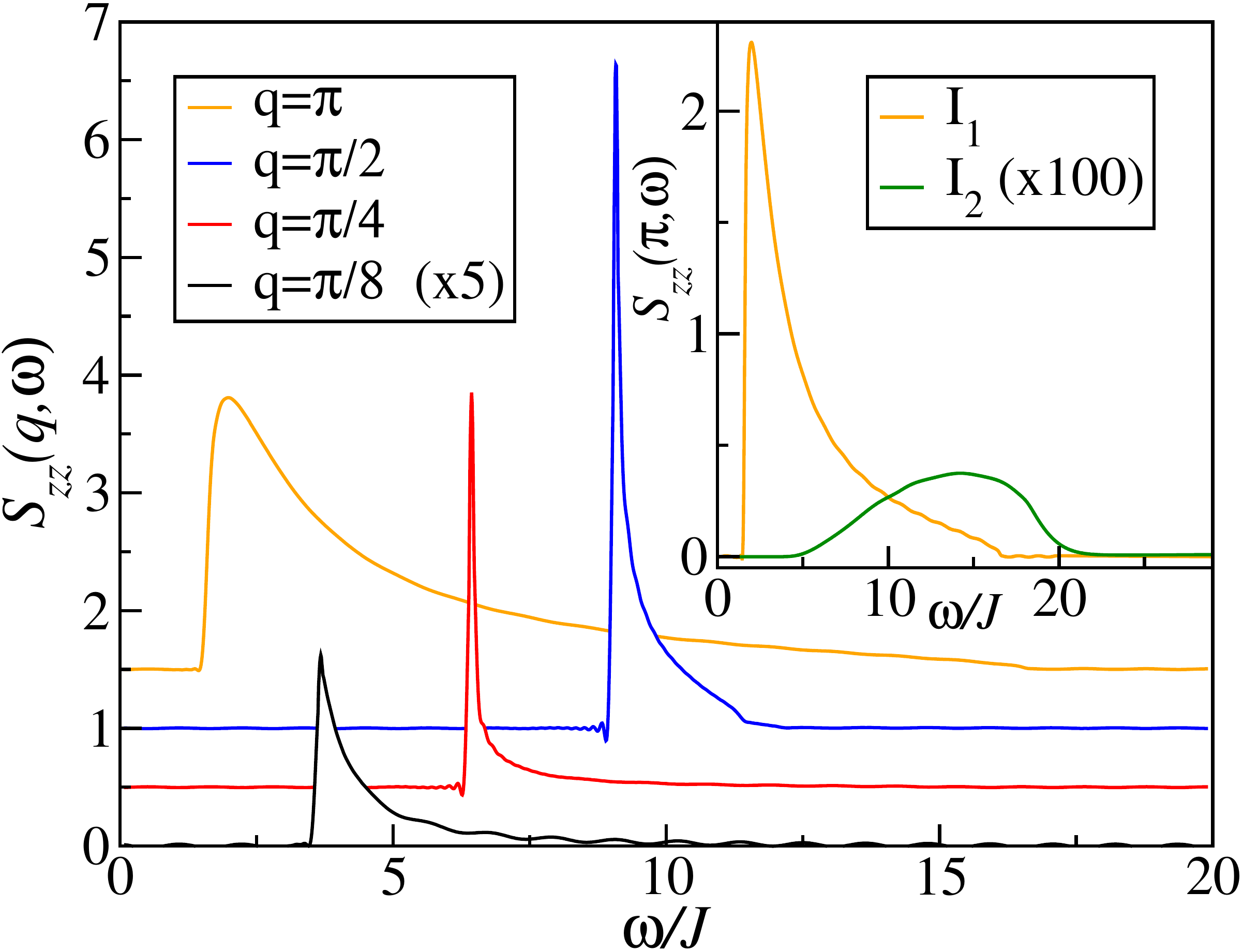} 
 \caption{$S_{zz}(q,\omega)$ with $\Delta=2$ for various wave numbers
 $q$. Note that subsequent curves are shifted vertically by
 $0.5$. Inset: The $\ell=1$ contribution is non-zero only within the
 two-spinon continuum $[\omega_{\rm low},\omega_{\rm up}]$ while
 $\ell=2$ excitations contribute also for $\omega>\omega_{\rm up}$.}
\label{fig:Skw} 
\end{figure}
%

Finally, as a first application of \eqref{Azz_formula}, we evaluate
the longitudinal DSF $S_{zz}(q,\omega)$, which is directly measurable
in neutron scattering experiments. In contrast to $S_{+-}$
\cite{BKM98}, the evaluation of $S_{zz}$ in the massive regime is
technically difficult and the two-spinon result within the VOA
has only recently been reported \cite{Castillo2020}. On the other hand,
\eqref{ffexpansion} and \eqref{Azz_formula} allow to obtain the
dynamical correlations for large $m$ and $t$, and we can readily
perform a numerical Fourier transform. We subtract the contribution of
the staggered magnetization and include both $\ell=1$ and $\ell=2$
excitations. The case $\Delta=2$ is plotted in Fig.~\ref{fig:Skw}
showing that the lineshapes as well as the weights for small $q$ are
well resolved. We have checked that the sum rules
\cite{Mueller82} are satisfied with good accuracy. The $\ell=1$ 
excitations are constraint to the spinon energy band with lower and
upper boundaries \cite{BKM98} $\omega_{\rm low} \sim 9.06 J,\,
\omega_{\rm up} \sim11.76 J
\,(q=\pi/2)$ and $\omega_{\rm low} \sim 1.56 J,\, \omega_{\rm up}
\sim16.56 J \,(q=\pi)$, respectively. Higher $\ell$ excitations lead to a 
high-frequency tail which becomes more prominent for $\Delta\to 1$
\cite{PereiraSirker}. For larger $\Delta$, the peaks will shift to
larger $\omega$ and have smaller amplitudes. For $T \ll 1$, the line
shape only weakly depends on the magnetic field if it is smaller than
the lower critical field, in sharp contrast to the massless case
($|\Delta|<1$).


To summarize, we have presented a closed-form expression, incorporating 
all orders of particle-hole excitations, for the dynamical longitudinal
correlation functions of the massive XXZ chain. This result opens up a
new avenue to understand the dynamical response of strongly
interacting quantum systems and is directly relevant for recent
experiments on cold atomic gases. It does also provide a
benchmark for the development of numerical algorithms including, for
example, recent attempts to learn quantum dynamics using neural
networks \cite{TroyerScience}. We note, furthermore, that the isotropic
Heisenberg model, $\Delta = 1$ in Eq.~(\ref{hamiltonian}), is immediately
obtained by rescaling all rapidities $x_j, y_{j}$ by $\gamma x_j,
\gamma y_{j}$ and by taking $\gamma \rightarrow 0$.  
As all  $I_{\ell}$ contribute with equal weight to the long-time
asymptotic behavior,  the explicit formula (\ref{Azz_formula}) will be
an indispensable tool to study this limit in detail.
Extensions to finite temperatures are promising and explicit
expressions are already available from thermal form factor expansions
if $T/J \ll 1$ \cite{BGKS20app}. Further progress in dealing with the
kinematic poles along the lines of Ref.~\cite{GKSS19} is expected
and will make it possible to access higher temperatures. A detailed
analysis of the dynamical structure factor will also be an important
subject of future studies.

\acknowledgments
We thank Andreas Kl{\"u}mper for his continuous encouragement. 
 C.B., F.G., and J.Si.~acknowledge financial support by the German
 Research Council (DFG) in the framework of the research unit FOR
 2316. K.K.K. is supported by CNRS Grant PICS07877. J.Si.~acknowledges
 support by the Natural Sciences and Engineering Research Council
 (NSERC, Canada).  J.Su.~is supported by JSPS KAKENHI Grant numbers
 18K03452, 18H01141.

%

%

\pagebreak

\onecolumngrid
\setcounter{equation}{0}
\setcounter{figure}{0}
\setcounter{table}{0}
\setcounter{page}{1}
\renewcommand{\theequation}{S\arabic{equation}}
\renewcommand{\thefigure}{S\arabic{figure}}
\renewcommand{\bibnumfmt}[1]{[S#1]}
\renewcommand{\citenumfont}[1]{S#1}

\begin{center}
  \textbf{\large Supplemental Material to\\
{``Exact real-time longitudinal correlation functions of the massive XXZ chain"}}\\[.2cm]
 Constantin Babenko, Frank G\"ohmann$^{1}$,  Karol K. Kozlowski$^{2}$, Jesko Sirker$^3$,  Junji Suzuki$^4$\\[.1cm]
  {\itshape ${}^1$Fakult\"at f\"ur Mathematik und Naturwissenschaften,
Bergische Universit\"at Wuppertal, 42097 Wuppertal, Germany\\
  ${}^2$Univ Lyon, ENS de Lyon, Univ Claude Bernard, CNRS,
Laboratoire de Physique, F-69342 Lyon, France\\
  ${}^3$Department of Physics \& Astronomy, University of Manitoba,
Winnipeg, Manitoba, Canada R3T 2N2\\
  ${}^4$Department of Physics, Faculty of Science,
Shizuoka University, Ohya 836, Suruga, Shizuoka, Japan\\}
(Dated: \today)\\[1cm]
\end{center}

\onecolumngrid
In this supplement we provide some background material and a sketch of
the derivation of our main result, Eq.\ (10) of the main text. The
starting point will be our recent work \cite{BGKS20appSM}.

\section{List of special functions}

As the main result necessarily includes many special functions,
we find it convenient to summarize their definitions.
The basic ingredients
of their definitions are the (multi-) $q$-Pochhammer symbols which,
for $|q_j| < 1$ and $a \in {\mathbb C}$, are defined as
\begin{equation}
     (a;q_1, \dots, q_p)_{\infty} =
        \prod_{n_1, \dots, n_p = 0}^\infty (1 - a q_1^{n_1} \dots q_p^{n_p}) .
\end{equation}
We will often drop $\infty$ in the subscript.
Based on this definition we introduce the $q$-gamma and $q$-Barnes
functions $\Gamma_q$ and $G_q$,
\begin{equation} \label{defgammaqgq}
     \Gamma_q (x) = (1 - q)^{1 - x} \frac{(q;q)}{(q^x;q)} , \quad
     G_q (x) = (1 - q)^{- \frac12 (1 - x)(2 - x)} (q;q)^{x - 1}
               \frac{(q^x;q,q)}{(q;q,q)} .
\end{equation}
All functions needed for the presentation of the final result of
\cite{BGKS20appSM} belong to the above $q$-gamma family.

We follow the definition of theta funtions in \cite{WhWa63ch21SM}.
Their product forms, e.g., read
\begin{align}\label{deftheta}
\vartheta_1(x|q)&= 2q^{\frac{1}{4}} \sin(x) \prod_{n=1}^{\infty} (1-q^{2n})(1-2q^{2n}\cos(2x) +q^{4n}) , &
\vartheta_3(x|q)&= \prod_{n=1}^{\infty} (1-q^{2n})(1+2q^{2n-1}\cos(2x) +q^{4n-2}) , \nonumber\\
\vartheta_2(x|q)&= 2q^{\frac{1}{4}} \cos(x) \prod_{n=1}^{\infty} (1-q^{2n})(1+2q^{2n}\cos(2x) +q^{4n}) , &
\vartheta_4(x|q)&= \prod_{n=1}^{\infty} (1-q^{2n})(1-2q^{2n-1}\cos(2x) +q^{4n-2}).
\end{align}
Recall that the other theta functions are related to, say, $\vartheta_4$ by shifts
of the arguments \cite{WhWa63ch21SM}.

The $q$-gamma and theta functions are related through the second
functional equation of the $q$-gamma function which may be written as
\begin{equation} \label{scdgammafuneq}
     \frac{\vartheta_4 (x|q)}{\vartheta_4 (0|q)} =
        \frac{\Gamma_{q^2}^2 \bigl({\frac12}\bigr)}
	     {\Gamma_{q^2} \bigl({\frac12 - \frac{i x}{\gamma}}\bigr)
	      \Gamma_{q^2} \bigl({\frac12 + \frac{i x}{\gamma}}\bigr)} .
\end{equation}

In our derivation of the explicit formula for the amplitudes, we shall
encounter the basic hypergeometric series \cite{GaRa04SM}, defined by
\begin{equation}\label{defHyperG}
     \phantom{}_r \Phi_s
        \left(
	\begin{array}{@{}r}
	   a_1, a_2, \cdots, a_r \\
	   b_1, b_2, \cdots, b_s
        \end{array}
        ;q,z \right)  
     = \sum_{n=0}^{\infty} \frac{(a_1, \cdots, a_r;q)_n}{(b_1, \cdots, b_s,q;q)_n} 
          \Bigl((-1)^n q^{\frac{n(n-1)}{2}}\Bigr)^{s+1-r} z^n,
\end{equation}
where we have used the following standard notation for $q$-Pochhammer symbols,
\begin{equation}
     (a;q)_l = \prod_{n=0}^{l-1} (1 - a q^n) , \quad
     (a_1, a_2,\cdots, a_k;q)_l = (a_1;q)_l   (a_2;q)_l \cdots  (a_k;q)_l .
\end{equation}

By definition $ \phantom{}_r \Phi_s$ is invariant under permutations of the
elements of the sets $\{a_j\}_{j=1}^r$ and $\{b_j\}_{j=1}^s$. In case that
they can be split into disjoint subsets, say, $\{a_j\}_{j=1}^r =
\{a^{(1)}_j\}_{j=1}^{l_1} \cup \{a^{(2)}_j\}_{j=1}^{l_2} \cup \cdots$,
in place  of (\ref{defHyperG}), we use a short-hand representation as employed
in $\Phi_1$ and $\Phi_2$ in the main text, 
 \begin{equation}
     \phantom{}_r \Phi_s
        \left(
	\begin{array}{@{}r}
	   \{a^{(1)}_j\}_{j=1}^{l_1}, \{a^{(2)}_j\}_{j=1}^{l_2},\cdots   \\
	     \{b^{(1)}_j\}_{j=1}^{l'_1} , \{b^{(2)}_j\}_{j=1}^{l'_2},\cdots
        \end{array}
        ;q,z \right).  
 \end{equation}

\section{QTM approach to dynamical correlation functions}
Following the general strategy of the classical-to-quantum correspondence,
we represent the correlation functions of the XXZ spin chain in terms of the
correlation functions of an inhomogeneous six-vertex model. The vertical
direction in the vertex model corresponds to discretized time and inverse
temperature. We use a discretization scheme with $2N + 2$ time and temperature
slices proposed in \cite{Sakai07SM} and combine it with a thermal form
factor expansion \cite{DGK13aSM,GKKKS17SM}.
We refer to  $N$  as the Trotter number.
It should be sent to infinity at the end of calculation (the Trotter limit).
This is a necessary procedure in order to recover
the original continuous time quantum system at finite temperature $T$. The
column-to-column transfer matrix acting on $2N+2$ sites is called the
(dynamical) quantum transfer matrix (QTM).

For the formal definition of the inhomogeneous six-vertex model we
have to introduce its  weights. They are encoded in the $R$-matrix
\begin{equation} \label{rmatrix}
     \begin{array}{cc}
     R(\lambda,\mu) = \begin{pmatrix}
                  1 & 0 & 0 & 0 \\
		  0 & b(\lambda,\mu) & c(\lambda,\mu) & 0 \\
		  0 & c(\lambda,\mu) & b(\lambda,\mu) & 0 \\
		  0 & 0 & 0 & 1
		 \end{pmatrix} , &
     \begin{array}{c}
     b(\lambda, \mu) = \frac{\sin(\mu - \lambda)}{\sin(\mu - \lambda + i \g)} \\[2ex]
     c(\lambda, \mu) = \frac{\sin(i \g)}{\sin(\mu - \lambda + i \g)}
    \end{array}
    \end{array} ,
\end{equation}
where $\gamma > 0$ is the same parameter as in $\Delta = \cosh(\gamma)$
and in $q = {\rm e}^{- \gamma}$.

Using (\ref{rmatrix}) we define a staggered, twisted and inhomogeneous
monodromy matrix acting on `vertical spaces' with `site indices'
$1, \dots, 2N + 2$, and on a `horizontal space' indexed $a$,
\begin{equation} \label{stagmon}
     T_a (\lambda|h) = {\rm e}^{h \s_a^z/2T}
        R_{2N+2, a}^{t_1} (\xi_{2N+2},\lambda + i \gamma/2)
	R_{a, 2N+1} (\lambda + i \gamma/2, \xi_{2N+1}) \dots
	R_{2, a}^{t_1} (\xi_2, \lambda) R_{a, 1} (\lambda + i \gamma/2, \xi_1) .
\end{equation}
The superscript $t_1$ denotes transposition with respect to the first
space on which $R$ is acting, and $\xi_1, \dots, \xi_{2N+2}$ are $2N+2$ complex
`inhomogeneity parameters'. 
We fix these parameters to the values \cite{GKKKS17SM,Sakai07SM}
\begin{equation} \label{trotterdecomp}
    \xi_{2j-1} = - \xi_{2j} =
                  \begin{cases}
		     - \frac{i t}{\kappa N} & j = 1, \dots, \frac{N}{2} \\[.5ex]
		     \epsilon & j = \frac{N}{2} + 1 \\[1ex]
		     \frac{i t + 1/T}{\kappa N} & j = \frac{N}{2} + 2, \dots, N+1 ,
                  \end{cases}
\end{equation}
where $1/\kappa = - 2 i J \sinh(\gamma)$ and $\epsilon \ne 0$ is a
regularization parameter.

By construction, the monodromy matrix $T_a (\lambda|h)$ can be seen
as a $2 \times 2$ matrix in space $a$, corresponding to a lattice
site in the spin chain, with entries $A(\lambda|h), \dots,
D(\lambda|h)$ that are operators acting on a $2^{2N+2}$ dimensional
auxiliary space,
\begin{equation}
T_a(\lambda|h) = 
\begin{pmatrix}
A(\lambda|h) & B (\lambda|h) \\
C(\lambda|h)&  D(\lambda|h) 
\end{pmatrix}_a .
\end{equation}

Thanks to the solution of the inverse problem, any local
spin operators in dynamical correlation functions
are expressible in terms of these operators.
The QTM $\tau(\lambda|h)$ and its twisted version
$\tau(\lambda|h')$ are defined by
\begin{equation}
     \tau(\lambda|h) ={\text tr}_a T_a(\lambda|h) = A(\lambda|h) + D(\lambda|h), \qquad
     \tau(\lambda|h') = {\rm e}^{\alpha \gamma} A(\lambda|h)
                        + {\rm e}^{-\alpha \gamma} D(\lambda|h),
\end{equation}
where the twist angle $\alpha$ is absorbed into the shift,  $\frac{h'}{2T}
=\frac{h}{2T}+\alpha \gamma$, see (\ref{stagmon}).

We denote the eigenvalues and eigenstates
of the QTM by $\Lambda_n (\lambda|h)$ and $|n, h\rangle$. Since
$\tau(\lambda|h)$ is non-Hermitian, its eigenvalues are generally not
all real. Still, there is a unique eigenstate $|0, h\rangle$, whose
eigenvalue is real and of largest modulus. We call this state the
dominant state. For brevity we shall suppress the suffix `0' and
write $\Lambda(\lambda|h)$ and $|h\rangle$ in this case.

Let us define $\Sigma^z(\lambda|h)$, which is related to the spin operator
$\sigma^z$ and  has a simple relation with $\tau(\lambda|h,\alpha) $, 
\begin{equation}\label{defSigmaz}
     \Sigma^z(\lambda|h) = A(\lambda|h) - D(\lambda|h)
        = \left. \partial_{\alpha \gamma} \tau(\lambda|h')  \right|_{\alpha=0}.
\end{equation}
We introduce an important object by  using $\Sigma^z(\lambda|h)$,  
\begin{equation} \label{defazz}
     A_n^{zz} (h) = \frac{\langle h|\Sigma^z (-i \gamma/2|h)|n, h \rangle
                          \langle n, h|\Sigma^z (-i \gamma/2|h)|h \rangle}
	                 {\Lambda_n (-i \gamma/2|h) \langle h|h\rangle
			  \Lambda (-i \gamma/2|h) \langle n, h|n, h\rangle}, 
\end{equation}
which will turn into ${\cal A}_{zz}$  in the end.

The longitudinal correlation function can then be represented as \cite{GKKKS17SM}
\begin{align} \label{deflongffseries}
     \bigl\langle \s_1^z(t) \s_{m+1}^z\bigr \rangle_T =
        \lim_{N \rightarrow \infty} \lim_{\epsilon \rightarrow 0}
	\sum_n A_n^{zz} (h) 
       \rho_n (- i \g/2|h)^m
               \biggl(\frac{\rho_n (- i \g/2 - i t/\kappa N|h)}
	                   {\rho_n (- i \g/2 + i t/\kappa N|h)}\biggr)^\frac{N}{2} ,
\end{align}
where $\rho_n(\la|h)= \La_n(\la|h)/ \La(\la|h)$.

By means of (\ref{defSigmaz}) one can rewrite (\ref{defazz}) as 
\begin{equation} \label{azzagen}
     A_n^{zz} (h) =
             \left.   \delta_{n,0} \biggl(\frac{\partial_{\alpha \gamma} \Lambda (- i \g/2|h)}
                                         {\Lambda (- i \g/2|h)} \biggr)^2   \right|_{\alpha=0} +
                \frac{1}{2} \bigl( \partial^2_{\alpha \gamma}  A_n(h,h')\bigr)_{\alpha=0}
                \biggl[ \rho_n (- i \g/2|h) - 2 +
		    \rho_n (- i \g/2|h)^{-1} \biggr]
\end{equation}
where we have introduced
\begin{equation} \label{genfunamps}
    A_n(h,h')= \frac{\langle h|n, h' \rangle  \langle n, h'|h \rangle}
                    {\langle h|h\rangle\langle n, h|n, h\rangle}
\end{equation}
and also used that $A_n(h,h)=0$ if $n \ne 0$.

\section{Summation, Trotter limit, and zero temperature limit}
Equation (\ref{deflongffseries}) gives us access to the longitudinal
two-point function at any finite temperature. For the evaluation of 
its right hand side we have to proceed in three steps. 1.) Evaluate
the amplitudes $A_n(h,h')$ and eigenvalue ratios $\rho_n(\cdot|h)$ at
finite Trotter number for all excited states. 2.) Sum up the series.
3.) Perform the Trotter limit.

A classification of all excited states at sufficiently low temperatures
was achieved in \cite{DGKS15bSM}. If, at finite magnetic field $h$,
the temperature is low enough, all excited states can be described
as particle-hole excitations.
 All string excitations disappear from
the spectrum. This made it possible to obtain all eigenvalue ratios
$\rho_n (\cdot|h)$ in closed form in this limit. In our recent work
\cite{BGKS20appSM} we attempted to use the same classification of
the eigenstates to obtain explicit expressions for the amplitudes
$A_n(h,h')$ defined in (\ref{genfunamps}). Using Slavnov's scalar
product formula \cite{Slavnov89SM} the right hand side is expressed
in terms of the sets of Bethe roots of the dominant and excited states
which parameterize certain `Slavnov determinants'. 
Taking the Trotter
limit of these determinants is not immediate and not a unique procedure.
In  the previous work \cite{DGKS16bSM} we obtained expressions for the
amplitudes involving Fredholm determinants of certain integral
operators that were valid in the low-$T$ limit.
 In our most recent work
\cite{BGKS20appSM}, inspired by \cite{KiKu19SM}, we
obtained expressions that involved only determinants of finite matrices
at low $T$. Here we only summarize the final result of \cite{BGKS20appSM}
and refer to the paper for more details. Starting from the final result
of \cite{BGKS20appSM} we can derive the fully explicit formula (10), 
reported in the Letter.

We  first introduce the functions that characterize the amplitude in the Trotter limit and
at $T = 0$. These are the bare energy
\begin{equation}
     e(\lambda) = \cot(\lambda) - \cot(\lambda - i \gamma) \,,\quad    
\end{equation}
a function with a complex parameter $\alpha$
\begin{equation} \label{defgalpha}
     g_\alpha (\lambda, \mu) =
        \frac{\vartheta_1'(0|q) \vartheta_2 (\mu - \lambda - \alpha|q)}
	     {\vartheta_2 (\alpha|q) \vartheta_1 (\mu - \lambda|q)} ,
\end{equation}
 the `averaged shift function'
\begin{equation} \label{defsigma0}
     \Sigma_0 = \frac12 \biggl( - k \pi + i \alpha \gamma
                                + \sum_{j=1}^\ell (y_j - x_j)\biggr) ,
\end{equation}
where the $y_j$ and $x_j$ are the particle and hole rapidities
introduced in the main text, and a pair of `weight functions'
defined for $\sigma = \pm$ as
\begin{equation} \label{defphpm}
     \Phi^{(\sigma)} (\lambda) =
        {\rm e}^{\sigma i \Sigma_0}
	\prod_{j = 1}^\ell
	\frac
	{\Gamma_{q^4} \Bigl(\frac12 - \frac{\sigma i (\lambda - x_j)}{2 \gamma}\Bigr)
	 \Gamma_{q^4} \Bigl(1 + \frac{\sigma i (\lambda - x_j)}{2 \gamma}\Bigr)}
	{\Gamma_{q^4} \Bigl(\frac12 - \frac{\sigma i (\lambda - y_j)}{2 \gamma}\Bigr)
	 \Gamma_{q^4} \Bigl(1 + \frac{\sigma i (\lambda - y_j)}{2 \gamma}\Bigr)}.
\end{equation}
They all enter in the definition of the key ingredients, $\Omega$ and
$\overline \Omega$,
\begin{subequations}
\begin{align}
     \Om (\la, \mu) 
        &= g_{\Sigma_0} (\la, \mu)
	  - \int_{- \pi/2}^{\pi/2} \frac{d z}{2 \pi i}
	                         \frac{\Phi^{(-)} (\mu)}{\Phi^{(+)} (z)} \,
				 g_{\Sigma_0} (\la, z) e(z - \mu),      \label{OmegaInt} \\[1ex]
     \overline{\Om} (\la,\mu) 
        &= g_{- \Sigma_0} (\la, \mu)
	  - \int_{- \pi/2}^{\pi/2} \frac{d z}{2 \pi i} \:
	                         \frac{\Phi^{(-)} (z)}{\Phi^{(+)} (\mu)} \,
				 g_{- \Sigma_0} (\la, z) e(\mu - z) .    \label{OmegabarInt} 
\end{align}
\end{subequations}

Finally the thermal form factor expansion of the longitudinal correlation
function as obtained in \cite{BGKS20appSM} reads 
\begin{multline}
     \langle \sigma_1^z (t) \sigma_{1+m}^z (0)\rangle
    =
     (-1)^m \frac{\vartheta_1'(0|q)^2}{\vartheta_2 (0|q)^2} \\ +
     \sum_{k=0,1} (-1)^{km} 
     \sum_{\ell=1}^{\infty} 
     \frac{1}{({\ell}!)^2} 
     \int_{C_-} \frac{d^{\ell} x}{(2\pi )^{\ell}} \int_{C_+} \frac{d^{\ell} y}{(2\pi )^{\ell}}
     {\cal A}_{zz}(\{x\},\{y\}|k)
     {\rm e}^{-i \sum_{j=1}^{\ell} \bigl( p(x_j) m -\varepsilon(x_j)t \bigr)}
     {\rm e}^{i \sum_{j=1}^{\ell} \bigl( p(y_j) m -\varepsilon(y_j)t \bigr)} ,
\end{multline}
where the amplitude  ${\cal A}_{zz}(\{x\},\{y\}|k)$ takes the form
\begin{multline} \label{amplitude}
     {\cal A}_{zz}(\{x\},\{y\}|k)= 
     (-1)^{\ell} 4 \sin^2\biggl( \frac{\pi k}{2}+ \sum_{j=1}^{\ell} 
      \frac{p(y_j) -p(x_j)}{2}  \biggr)
     \frac{\vartheta^2_2(\Sigma_0|q)}{\vartheta^2_2(0|q)}
	  \prod_{i,j=1}^{\ell}
	  \frac{\psi(x_i - x_j) \psi(y_i - y_j) }{\psi(x_i-y_j) \psi(y_j-x_i)} \\ \times
     \left. \frac{\partial {\rm det} \,\Omega (x_i, y_j) }{\partial i\alpha \gamma} 
     \frac{\partial {\rm det} \,\overline{\Omega} (x_i, y_j)}
          {\partial i\alpha \gamma}  \right|_{\alpha=0} . 
\end{multline}

\section{The derivation of the explicit formula for the amplitude}

We have to take the $\alpha$-derivatives explictly in (\ref{amplitude}) to
obtain the amplitudes ${\cal A}_{zz}$ that determine the physical correlation function.
Each matrix element $\Om (x_i,y_j) $ is calculated from (\ref{OmegaInt}) and (\ref{OmegabarInt}), 
and the only $\alpha$ dependency appears through $\Sigma_0$. It is then easy to check
that the column vectors or the row vectors of the matrix $\Om$ are not simply parallel.
As the existence of a null vector  of  $\Om$ is proven \cite{BGKS20appSM}, we can safely perform 
a Taylor expansion with respect to $\alpha$. It results, however, in a sum over
many terms which is inconvenient for practical use. Thus, the first order zeros in
$\alpha$ are better factored out explicitly from ${\rm det}\, \Omega$ and ${\rm det}\,
\overline{\Omega}$.

The integral in (\ref{OmegaInt}) can be calculated by means of the residue
theorem as the integrand is $\pi$-periodic and meromorphic. It has infinite
series of simple poles in the upper and in the lower half plane. We obtain
two different representations of $\Om(x_i,y_j)$ by closing the integration
contour in the upper or in the lower half plane. These representations can
be neatly written in terms of special $\alpha$-dependent basic hypergeometric
series,
\begin{subequations}
\begin{align}
     \Phi_1 \bigl(P_j, \{P_l\}_{ l \ne j}, \{H_{r}\},\alpha\bigr)
        & = \phantom{}_{2\ell} \Phi_{2\ell-1}
	    \left(
	    \begin{array}{@{}r}
	       q^{-2}, \{ q^2  \frac{P_j} {P_{l}} \}_{l \ne j},
	       \{\frac{P_j} {H_{r}}\}_{r=1}^{\ell}  \\
               \{\frac{P_j} {P_{l}} \}_{l \ne j}, 
	       \{q^2\frac{P_j} {H_{r}} \}_{r=1}^{\ell}
	    \end{array};
	    q^4, q^4 {\rm e}^{-2\alpha \gamma}\right), \label{defPhi1} \\[1ex]
     \Phi_2 \bigl(P_j, P_i, \{P_l\}_{ l \ne i,j}, \{H_{r}\},\alpha\bigr)
        & = \phantom{}_{2\ell} \Phi_{2\ell-1}
            \left(
	    \begin{array}{@{}r}
	       q^{6}, \{q^6 \frac{P_i} {P_{l}}\}_{l \ne i,j}, 
	       q^2 \frac{P_i} {P_j}, \{  q^4\frac{P_i} {H_{r}} \}_{r=1}^{\ell} \\
               \{q^4 \frac{P_i} {P_{l}}\}_{l \ne i,j}, q^8 \frac{P_i} {P_j}, 
	       \{q^6\frac{P_i} {H_{r}}\}_{r=1}^{\ell}
	    \end{array};
	    q^4, q^4 {\rm e}^{-2\alpha \gamma}\right), \label{defPhi2} \\[1ex]
     \Phi_3 \bigl(P_j, H_i, \{P_l\}_{ l \ne j}, \{H_{r}\}_{r \ne i} ,\alpha\bigr) 
        & = \phantom{}_{2\ell} \Phi_{2\ell-1}
	    \left(
	    \begin{array}{@{}r}
	       q^{2}, \frac{H_i}{P_j}, \{q^4\frac{H_i}{P_{l}}\}_{l \ne j},
	       \{q^2 \frac{H_i}{H_{r}}\}_{r \ne i} \\
	       q^6 \frac{H_i}{P_j}, \{q^2\frac{H_i}{P_{l}}\}_{l \ne j},
	       \{q^4 \frac{H_i}{H_{r}} \}_{r \ne i}
	    \end{array};
	q^4, q^4 {\rm e}^{-2\alpha \gamma} \right). \label{defPhi3}
\end{align}
\end{subequations}
Here the variable set $\{*\}$ means that the function is invariant
under permutations among the members of $*$. We will drop the symmetric
arguments and use the abbreviations $\Phi_1(P_j,\alpha)$,
$\Phi_2(P_j, P_i, \alpha)$ or $\Phi_3 (P_j, H_i, \alpha)$. We further set 
\begin{equation}\label{Psi2}
     \Psi_2(P_j, P_i,\alpha)
        = {\rm e}^{-2\alpha \gamma} r_{\ell} (P_j,P_i) \Phi_2(P_j, P_i,\alpha), 
\end{equation}
where $r_{\ell}$ is defined in Eq.\ (7) of the main text.

The calculation of the sum of the residues is straightforward but rather
cumbersome. It finally results in
\begin{align}
     \Om(x_j, y_i) & = - \frac{\Phi^{(-)} (y_i)}{\Phi^{(+)} (y_i)} 
                       g_{\Sigma_0} (x_j, y_i) \Phi_1 (P_i, \alpha)
                     + \sum_{\substack{l = 1\\l \ne i}}^\ell
                       \frac{\Phi^{(-)} (y_i)}{\Phi^{(+)} (y_l)} 
                       g_{\Sigma_0} (x_j, y_l) \Psi_2 (P_i, P_l, \alpha)
		       \notag \\[-4ex] & \mspace{489.mu}
                     + \frac{\Phi^{(-)} (y_i)}{\Phi^{(-)} (x_j)} e(y_i - x_j)
		       \Phi_3 (P_i, H_j, \alpha), \notag \\[1ex]
                   & = g_{\Sigma_0} (x_j, y_i) \overline{\Phi}_1 (P_i, \alpha)
                     - \sum_{\substack{l = 1\\l \ne i}}^\ell
                       \frac{\Phi^{(-)} (y_i)}{\Phi^{(-)} (y_l)} 
                       g_{\Sigma_0} (x_j, y_l) \overline{\Psi}_2 (P_i, P_l, \alpha)
                     + \frac{\Phi^{(-)} (y_i)}{\Phi^{(+)} (x_j)} e(x_j - y_i)
		       \overline{\Phi}_3 (P_i, H_j, \alpha). \label{evalomega}
\end{align}
Slightly generalizing the definition in the main text the overbar means
here that $H_j \rightarrow 1/H_j$, $P_i \rightarrow 1/P_i$, and
$\alpha \rightarrow - \alpha$. The first equation is the result of 
closing the contour in the upper half plane, the second equation comes
from closing the contour in the lower half plane.

Neither of the two expressions (\ref{evalomega}) looks particularly useful, 
but they can be used to simplify one another. The key observation
is that they have a simple dependence on the index $k$. It enters $\Om$ only
through $\Sigma_0$ (cf.\ Eq.\ (\ref{defsigma0})). In order to make this
dependence explicit we introduce
\begin{equation}
     \sigma_0 = \Sigma_0 + \frac{\pi k}{2}
\end{equation}
which is independent of $k$. Then $\Phi^{(\pm)} (\lambda) = (-1)^k
{\rm e}^{\pm i \sigma_0} \phi_0^{(\pm)} (\lambda)$, where $\phi_0^{(\pm)}
(\lambda)$ is independent of $k$. The dependence of $g_{\Sigma_0}$ on $k$
is slightly more tricky to work out. Setting
\begin{equation}
     \theta_a(w,\sigma_0) = \frac{\vartheta_a(2\sigma_0 -w|q^2)} {\vartheta_a(w|q^2)}
\end{equation}
for $a = 1, 4$ we see that
\begin{equation}
     g_{\Sigma_0} (x,y) = - \theta_1' (2 \sigma_0, \sigma_0)
        \bigl(\theta_1 (y - x, \sigma_0) + (-1)^k \theta_4 (y - x, \sigma_0)\bigr).
\end{equation}

Altogether, using these simple formulas, the two expressions for $\Om$
take the form $\Om (x_j, y_i) = (-1)^k A + B = (-1)^k C + D$, where $A$,
$B$, $C$, $D$ are independent of $k$ and $A$, $B$ refer to the first
equation in (\ref{evalomega}), whereas $C$, $D$ refer to the second equation.
While $A$ and $D$ have forms that are convenient for factorization, 
$B$ and $C$ do not. Thanks to the equality, however, it follows that
$A = C$, $B = D$ and therefore $\Om (x_j, y_i) = (-1)^k A + D$. 
That is, we can represent $\Om$ entirely in terms of convenient expressions.
Spelling this out explicitly we obtain
\begin{multline}
     \Om (x_j, y_i) = - \sum_{l=1}^\ell \biggl\{\delta_{il}
                        \biggl[\overline{\Phi}_1 (P_i, \alpha) -
			       \frac{\Phi^{(-)} (y_i)}{\Phi^{(+)} (y_i)}
			       \Phi_1 (P_i, \alpha)\biggr] \\
        - (1 - \delta_{il})\biggl[\frac{\Phi^{(-)} (y_i)}{\Phi^{(-)} (y_l)}
	                       \overline{\Psi}_2 (P_i, P_l, \alpha) -
                               \frac{\Phi^{(-)} (y_i)}{\Phi^{(+)} (y_l)}
	                       \Psi_2 (P_i, P_l, \alpha)\biggr]\biggr\}
			       \theta_1' (2 \sigma_0, \sigma_0)
			       \theta_1 (y_l - x_j, \sigma_0)
\end{multline}
which is a matrix element of a product of two matrices. The factor
$(-1)^k$, that appears at an intermediate stage, is re-absorbed in
$\Phi^{(\pm)}$ here.

Thus,
\begin{equation}
     \det \Om (x_j, y_i) =
        (-1)^\ell \det \bigl\{\theta_1' (2 \sigma_0, \sigma_0)
	                      \theta_1 (y_l - x_j, \sigma_0)\bigr\}
        \det \{\mathcal{M} (\alpha)\},
\end{equation}
where
\begin{equation}
     \mathcal{M} (\alpha)_{i j} =
        \delta_{ij} \biggl[\overline{\Phi}_1 (P_j, \alpha) -
	                \frac{\Phi^{(-)} (y_j)}{\Phi^{(+)} (y_j)} \Phi_1 (P_j, \alpha)\biggr]
        - (1 - \delta_{ij})\biggl[\overline{\Psi}_2 (P_j, P_i, \alpha) -
                               \frac{\Phi^{(-)} (y_i)}{\Phi^{(+)} (y_i)}
	                       \Psi_2 (P_j, P_i, \alpha)\biggr].
\end{equation}
In the last step, we divide the $i$-th row by  $\Phi^{(-)} (y_i)$ and  multiply the $l$-th
column by $\Phi^{(-)} (y_l)$, which leaves the determinant invariant.

The first determinant is of elliptic Cauchy (or Frobenius) type and
can be calculated in product form,
\begin{equation}
     \det \bigl\{\theta_1'(2 \s_0, \s_0) \theta_1(y_i - x_j, \s_0)\bigr\} =
        \frac{\vartheta_1 (i \alpha \g|q^2)}
	     {\vartheta_1 (2 \s_0|q^2)} \bigl(\vartheta_1' (0|q^2)\bigr)^\ell \,
	\frac{\prod_{1 \le j < r \le \ell} \vartheta_1 (x_j - x_r|q^2) \vartheta_1 (y_r - y_j|q^2)}
	     {\prod_{j,r = 1}^\ell \vartheta_1 (x_j - y_r|q^2)}.
\end{equation}

Note that the prefactor is of order $\alpha$ here. This makes it easy
to calculate the $\alpha$-derivative of $\det \Om$ occurring on the
right hand side of (\ref{amplitude}),
\begin{equation}
     \frac{\partial \det \Om (x_j, y_i)}{\partial i \alpha \gamma}\biggr|_{\alpha = 0}
     = (-1)^\ell
        \frac{\vartheta_1' (0|q^2)\bigl(\vartheta_1' (0|q^2)\bigr)^\ell}
	     {\vartheta_1 (\sum_j (y_j - x_j)|q^2)} \,
	\frac{\prod_{1 \le j < r \le \ell} \vartheta_1 (x_j - x_r|q^2) \vartheta_1 (y_r - y_j|q^2)}
	     {\prod_{j,r = 1}^\ell \vartheta_1 (x_j - y_r|q^2)} \det \{ \mathcal{M} (0)\}.
\end{equation}
A similar formula is obtained for the $\alpha$-derivative of
$\det \overline{\Om}$ if we take into account that
\begin{equation} \label{eq:PH_Transform}
     \overline{\Omega}(y_j, x_i) =  
        - \left. \Omega(x_j,y_i)
	  \right |_{ x_j \rightarrow -y_j,  y_i \rightarrow -x_i, \alpha \rightarrow \alpha}.
\end{equation}
Inserting the results for the derivatives of $\det \Om$ and of
$\det \overline \Om$ back into (\ref{amplitude}) and using a few
standard theta function identities we arrive at Eqs.\ (5)-(11)
of the main text.

\bibliographystyle{amsplain}
\providecommand{\bysame}{\leavevmode\hbox to3em{\hrulefill}\thinspace}
\providecommand{\MR}{\relax\ifhmode\unskip\space\fi MR }
\providecommand{\MRhref}[2]{%
  \href{http://www.ams.org/mathscinet-getitem?mr=#1}{#2}
}
\providecommand{\href}[2]{#2}

\end{document}